\documentclass[aps,pra,preprint,amsmath, superscriptaddress, 11pt]{revtex4-2}
\usepackage{graphicx}
\usepackage{dcolumn}
\usepackage{bm}
\usepackage{mathrsfs}
\usepackage{subfigure}
\usepackage{natbib}
\usepackage{amssymb}
\usepackage{amsfonts}
\usepackage{multirow}
\usepackage{float}
\usepackage{color}
\usepackage{lipsum}
\usepackage{bbm}
\usepackage{caption}
\usepackage{hyperref}
\usepackage{latexsym}
\usepackage{upgreek}
\usepackage{amsmath}
\usepackage{ulem}

\DeclareMathAlphabet\mathbfcal{OMS}{cmsy}{b}{n}

\usepackage{changes}

  \colorlet{Changes@Color}{blue}

\begin{document}

\title{Yu-Shiba-Rusinov states in $s$-wave kagome superconductors: Self-consistent Bogoliubov-de Gennes calculations}

\author{Shuaibo Ding}
\affiliation{Zhejiang Key Laboratory of Quantum State Control and Optical Field Manipulation, Department of Physics, Zhejiang Sci-Tech University, 310018 Hangzhou, China}

\author{Yunfei Bai}
\affiliation{Zhejiang Key Laboratory of Quantum State Control and Optical Field Manipulation, Department of Physics, Zhejiang Sci-Tech University, 310018 Hangzhou, China}

\author{A. A. Bulekov}
\affiliation{HSE University, 101000 Moscow, Russia}

\author{Wenhui Zhang}
\affiliation{School of Electronic Engineering, Nanjing Xiaozhuang University, Nanjing, China}

\author{A. A. Shanenko}
\affiliation{HSE University, 101000 Moscow, Russia}

\author{Yajiang Chen}
\email[]{yjchen@zstu.edu.cn}
\affiliation{Zhejiang Key Laboratory of Quantum State Control and Optical Field Manipulation, Department of Physics, Zhejiang Sci-Tech University, 310018 Hangzhou, China}

\begin{abstract}
Significant research has recently been conducted into the Yu-Shiba-Rusinov (YSR) states in kagome superconductors through theoretical modeling and experimental investigations. However, additional efforts are still needed to further understand the local superconductivity near magnetic impurities in the kagome lattice and clarify how relevant quantities depend on the interaction strength $J$ between such impurities and electrons. In this study, we explore a self-consistent numerical solution of the Bogoliubov-de Gennes equations for an $s$-wave superconducting kagome model with a single classical magnetic impurity. Our study reveals that with increasing $J$, the local pair potential is systematically depressed in the vicinity of the impurity, similar to previous results obtained for the square and triangular lattices. Moreover, when further increasing $J$, the system undergoes a first-order phase transition with the appearance of stable and metastable states, reflecting the presence of the hysteresis loop in the pertinent quantities. As a consequence of this transition, the minimal energy of the stable YSR state is nonzero at any $J$, contrary to the expectations based on the assumption of a constant pair potential. A distinctive feature of the kagome lattice is that characteristics of the first-order transition are very sensitive to the position of the chemical potential within the kagome energy spectrum. 
\end{abstract}
\maketitle

Kagome metals and superconductors have been investigated intensively because of rich exotic properties induced by their geometric frustration and strongly correlated interactions. For instance, the theoretically predicted~\cite{guo2009, ko2009, kiesel2012, wang2013c} flat band (FB), van Hove singularity (VHS), Dirac point (DP), and multiband feature of the kagome-lattice energy spectrum have been experimentally verified in the high-quality kagome metals, e.g., Fe$_3$Sn$_2$~\cite{ye2018a}, AV$_3$Sb$_5$(A=K, Rb, Cs)~\cite{ortiz2019, ortiz2020, hu2022a}, YCr$_6$Ge$_6$~\cite{wang2020b}, and AV$_6$Sn$_6$(A=Y, Gd)~\cite{pokharel2021}. In addition, the angle-resolved photoemission spectroscopy~\cite{ortiz2020} and measurements of resistivity and magnetic susceptibility~\cite{chen2021c} have revealed multiple electron orders resulting from complex electron interactions in these materials, such as the topological ordering and charge-density waves. Moreover, unconventional superconductivity has been predicted~\cite{ko2009} and observed in CsV$_3$Sb$_5$~\cite{mielke2022}, including the nematic phase~\cite{nie2022}, pair-density waves~\cite{chen2021}, the M-like pressure-dependence of $T_c$~\cite{yu2021}, and the superconducting order parameter with possible two-band ($s$+$s$)-wave pairing~\cite{xu2021, gupta2022}, etc.

Recently, the Yu-Shiba-Rusinov (YSR) states~\cite{yu1965, shiba1968, rusinov1969a, balatsky2006}, as a promising candidate for constructing Majorana zero modes and qubits~\cite{Kitaev2001, morr2003,choy2011,nadj-perge2013,klinovaja2013}, have been studied in kagome superconductors both theoretically~\cite{basak2022} and experimentally~\cite{xu2021}. The YSR states are known to be bound in-gap quasiparticle states localized near magnetic impurities~\cite{yu1965, shiba1968, rusinov1969a}. Within the simplest scenario, when $J$ surpasses the critical value $J_c$, the energy of the YSR state crosses zero, leading to a localized pair breaking near the magnetic impurity. For the $s$-wave superconducting kagome model, the Bogoliubov-de Gennes (BdG) calculations~\cite{basak2022} have demonstrated that $J_c$ is very sensitive to 
the presence of the van Hove singularities (VHS), Dirac point (DP), and flat band (FB) in the kagome single-particle energy spectrum, which opens prospects of investigating the interplay between the YSR states and DOS singularities. Here it is worth noting that the system in question can be constructed by fabricating an artificial kagome lattice on top of an $s-$wave superconductor due to the proximity effect~\cite{lin2022, farinacci2023}. 

However, the BdG solution employed in Ref.~\onlinecite{basak2022} is not self-consistent, as it is based on the assumption of a uniform constant pair potential independent of $J$. In contrast, the self-consistent BdG calculations for the square and triangular lattices~\cite{salkola1997, glodzik2018} have demonstrated that increasing the coupling between an $s$-wave superconductor and a magnetic impurity results in a first-order phase transition with significant depression of the condensate near the impurity and the related $0$-$\pi$ phase shift (i.e., change of the sign) of the local pair potential at the impurity site. Such depression has recently been detected around a magnetic Cr cluster in CsV$_3$Sb$_5$ by scanning tunneling microscopy~\cite{xu2021}. The presence or absence of this particular type of first-order phase transition in kagome lattices has not been investigated so far, and our current study aims to fill this gap.

In the present work, by numerically solving the BdG equations in a self-consistent manner, we systematically investigate the properties of the YSR states in an $s-$wave superconducting kagome model. Our numerical results demonstrate that the system undergoes a first-order transition when increasing the superconductor-impurity coupling $J$. The related hysteresis loop of this transition appears in the interval $J_{\rm c1} \leq J \leq J_{\rm c2}$, where $J_{\rm c1},J_{\rm c2}$ correspond to the metastable zero-energy YSR states. However, the minimal YSR energy of the stable branch (i.e., the branch with the minimal free energy) is nonzero. In addition, the local pair potential of the stable branch exhibits a significant suppression near the impurity, with a jump at the transition point and the related $0$-$\pi$ phase shift, in agreement with the results reported previously for the square and triangular lattices~\cite{salkola1997, glodzik2018}. Characteristics of the first-order transition are essentially dependent on the chemical potential $\mu$ and, thus, sensitive to the VHS, DP, and FB of the kagome energy spectrum.

The starting point of our investigation is an $s-$wave superconducting kagome model with a single classical magnetic impurity. Its grand-canonical Hamiltonian is given by~\cite{Essler2005}
\begin{equation}\label{fullH}
    H=\sum_{ij\sigma}H_{ij\sigma} c_{i\sigma}^\dagger c_{j\sigma}-\frac{g}{2}\sum_{i\sigma\sigma'} c_{i\sigma}^\dagger c_{i\sigma'}^\dagger c_{i\sigma'} c_{i\sigma},
\end{equation}
where the indices $i=\{i_x, i_y\}$ and $j=\{j_x, j_y\}$ enumerate the sites of the two-dimensional kagome lattice; $c_{i\sigma}$ and $c^\dagger_{i\sigma}$ are the fermionic annihilation and creation operators for electrons at the site $i$ with the spin projection $\sigma=\pm1$ in units of $\hbar/2$; $g(>0)$ is the $s$-wave pair coupling; and $H_{ij\sigma}$ is the matrix element of the single-particle Hamiltonian. The latter, for the case of the nearest-neighbor approximation, reads $H_{ij\sigma}=-t\delta_{\langle ij \rangle}-\left[\mu-\left(K-\sigma J\right)\delta_{ii_0}\right]\delta_{ij}$~\cite{correction}, with $t$ the nearest-neighbor hoping parameter and $K$ the scattering potential coupling (nonmagnetic). The magnetic impurity is located at $i_0=\{i_{0x},\,i_{0y}\}$.

Within the mean-field approximation~\cite{degennes1964}, the Hamiltonian given by Eq.~(\ref{fullH}) acquires the form 
\begin{equation}\label{Heff}
H_{\rm eff} = \sum_{ij\sigma}H_{ij\sigma} c_{i\sigma}^\dagger c_{j\sigma} 
+\sum_i\left(\Delta_i c_{i\uparrow}^\dagger c_{i\downarrow}^\dagger+\Delta_i^* c_{i\downarrow} c_{i\uparrow}\right),
\end{equation}
with $\Delta_{i}$ being the pair potential. For simplicity, the Hartree-Fock potential~\cite{ketterson1999} has been ignored in Eq.~(\ref{Heff}) as its effect is only reduced to a shift in the chemical potential which does not change the general picture of the numerical results~\cite{chen2009, chen2012a, chen2014, yin2023, chen2024}. Using the Bogoliubov unitary transformation~\cite{basak2022} $c_{i\sigma}=\sum_{n}(u_{in\sigma}\gamma_{n\sigma}-\sigma v_{in\sigma}^*\gamma_{n\overline{\sigma}}^\dagger)$ with $\overline{\sigma}=-\sigma$, the effective Hamiltonian Eq.~(\ref{Heff}) is diagonalized as $H = \sum_{n\sigma} \varepsilon_{n\sigma} \gamma_{n\sigma}^\dagger\gamma_{n\sigma}+E_0$, where $\varepsilon_{n\sigma}$ is the quasiparticle energy, $u_{in\sigma}$ and $v_{in\sigma}$ are the spatial particle- and hole-like quasiparticle wavefunctions, $\gamma_{n\sigma}$ and $\gamma_{n\overline{\sigma} }^\dagger$ are the quasiparticle creation and annihilation operators satisfying the fermionic canonical commutation relations, and $E_0=\sum_i|\Delta_i|^2/g-\sum_{in\sigma}|v_{in\overline{\sigma}}|^2 \varepsilon_{n\sigma}$ is the ground state energy~\cite{ketterson1999}. The quasiparticle energy ($\varepsilon_{n\sigma}$) and the wave functions ($u_{in\sigma}$ and $v_{in\sigma}$) obey the Bogoliubov-de Gennes (BdG) equations
\begin{subequations}\label{bdg}
\begin{align}
 \sum_j H_{ij\sigma} u_{in\sigma} + \Delta_i v_{in\overline{\sigma}}  &= \varepsilon_{n\sigma} u_{in\sigma} \\
\Delta^*_i u_{in\sigma} - \sum_j H^*_{ij\overline{\sigma}} v_{jn\overline{\sigma}} &= \varepsilon_{n\sigma} v_{in\overline{\sigma}},
\end{align}
\end{subequations}
where solutions to the BdG equations are chosen so that to satisfy the normalization condition $\sum_i|u_{in\sigma}|^2 + |v_{in\overline{\sigma}}|^2 =1$ and the periodic boundary conditions for $u_{in\sigma}$ and $v_{in\sigma}$.

The BdG equations are solved together with the self-consistency relation of $\Delta_i$ given by
\begin{equation}\label{OP}
  \Delta_i = g \sum_{0\le\varepsilon_n\le\hbar\omega_D} u_{in\uparrow}v_{in\downarrow}^*(1-f_{n\uparrow}) - u_{in\downarrow}v^*_{in\uparrow}f_{n\downarrow},
\end{equation}
with the Fermi-Dirac quasiparticle distribution $f_{n\sigma} =f(\varepsilon_{n\sigma})$. The summation for $\Delta_{i}$ in Eq.~(\ref{OP}) includes only the quasiparticle states with positive energies inside the Debye energy ($\hbar\omega_D$) window, i.e., $0\leq\varepsilon_{n\sigma} \leq\hbar \omega_D$~\cite{degennes1964}. The obtained self-consistent solutions are used to calculate all the pertinent quantities, including the free energy $F$ and the spatial electron density $n_{ei}$~\cite{kosztin1998a}
\begin{subequations}\label{F_ne}
\begin{align}
F &=  \sum_{n\sigma} f_{n\sigma}\varepsilon_{n\sigma}+\sum_i|\Delta_i|^2/g-\sum_{in\sigma}|v_{in\overline{\sigma}}|^2\varepsilon_{n\sigma}, \label{F}\\
n_{ei} &=  \sum_{ n\sigma} |u_{in\sigma}|^2f_{n\sigma} + |v_{in\sigma}|^2(1-f_{n\overline{\sigma}}), \label{ne}
\end{align}
\end{subequations}
where $F$ is given for the zero temperature.

\begin{figure*}[ht]
\centering
\includegraphics[width=0.7\linewidth]{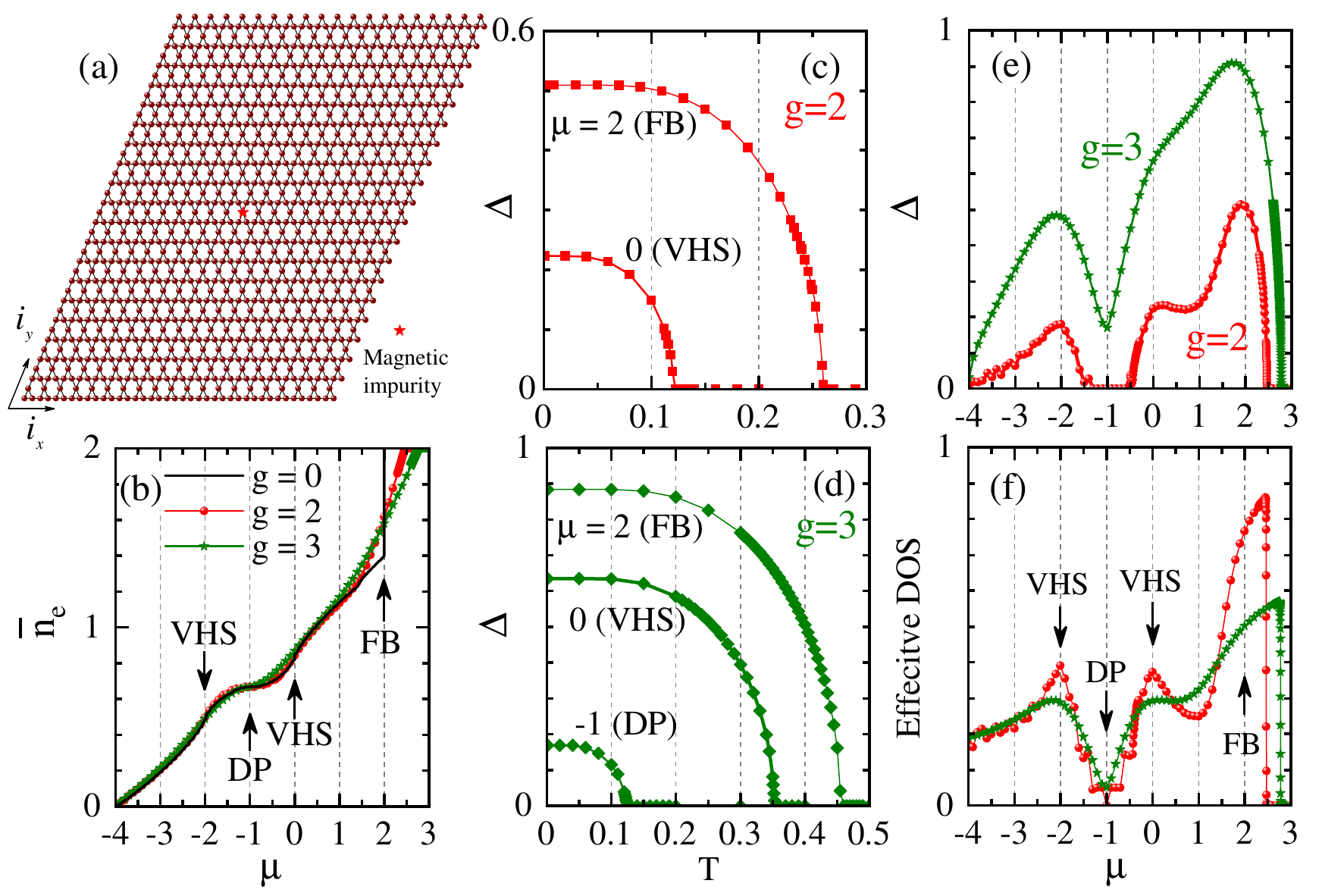}
\caption{(Color online) (a) Sketch of the kagome lattice used in our calculations and containing $20\times20$ primitive unit cells with a single magnetic impurity marked by the red star; (b) The self-consistent impurity-free filling level $\overline{n}_e$ as a function of the chemical potential $\mu$ for $g=0$ (black curve), $2$ (red spheres) and $3$ (green stars) at $T=0$, with highlighting the particular values of $\mu$ corresponding to the van Hove singularities (VHS), Dirac point (DP), and flat band (FB); (c, d) The self-consistent impurity-free $\Delta=\Delta_i$~(independent of $i$) versus $T$ for selected values of $\mu$ at $g=2$ and $3$; (e, f) $\Delta$ and the effective density of states $d \overline{n}_e/d \mu$ as functions of $\mu$ for $g=2$ (spheres) and $3$ (stars) at $T=0$ in the absence of magnetic impurities.}
\label{fig1}
\end{figure*}

The BdG equations given by Eq.~(\ref{bdg}) can be separated into the two sub-sets including the quasiparticle energies $\varepsilon_{n,\uparrow}$ and $\varepsilon_{n,\downarrow}$, respectively. These sub-sets are connected to one another by the relations $\varepsilon_{n\downarrow}=-\varepsilon_{n\uparrow}$, $u_{in\downarrow}= v_{in\downarrow}^*$, and $v_{in\uparrow}= -u_{in\uparrow}^*$. Thus, to simplify the numerical procedure, we solve only the sub-set of the BdG equations including $\varepsilon_{n\uparrow}$ and, then, the above relations are utilized to get the corresponding solutions for $\varepsilon_{n\downarrow}$. To check how the properties of the YSR states are affected by the proximity to the VHS, DP, and FB, we follow the choice of Ref.~\cite{basak2022} and consider the chemical potential $\mu$ as an external parameter, rather than the average electron filling level ($\overline{n}_{e}=\sum_i n_{ei}/N_{\rm tot}$, with $N_{\rm tot}$ being the total number of sites) as in Refs.~\cite{bai2023, bai2023a}. 

When $J$ is sufficiently small or large (i.e., beyond the hysteresis loop of the related first-order transition), numerical self-consistent BdG calculations are performed as follows. First, we take an arbitrary non-zero initial guess for $\Delta_i$. Second, utilizing this guess, we solve the BdG equations~(\ref{bdg}), obtaining the corresponding quasiparticle energies $\varepsilon_{n\sigma}$ and wavefunctions (i.e., $u_{in\sigma}$ and $v_{in\sigma}$). Third, using these energies and wave functions, we calculate a new pair potential from Eq.~(\ref{OP}) and repeat the second step until the numerical convergence of $\Delta_i$ is achieved.

\begin{figure*}[t]
\centering
\includegraphics[width=1\linewidth]{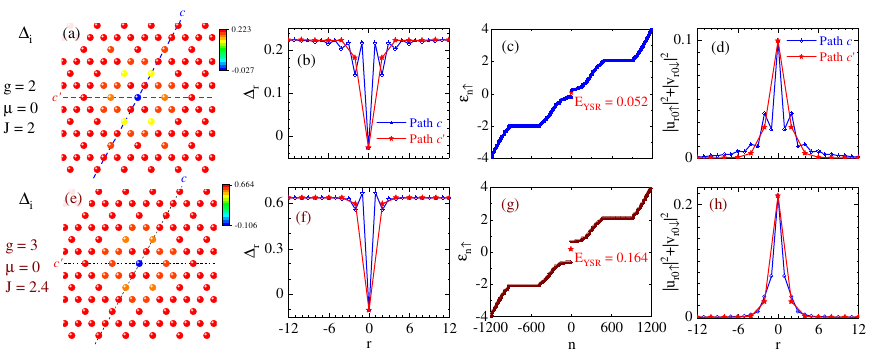}
\caption{(Color online) Self-consistent numerical results for $(g,\,J)=(2,\, 3)$ and $(3,\, 2.4)$ at $T=0$ and $\mu=0$ (VHS): (a,e) The contour plot of the pair potential $\Delta_i$, with $i=\{i_x,i_y\}$, near the magnetic impurity (the impurity is located at the crossing point of the lines $c$ and $c'$). (b,f) The spatial profile of the pair potential along the $c$ and $c'$ directions $\Delta_r$ versus the distance to the impurity $r$~(in units of $a$); (c,g) The quasiparticle energy spectra $\epsilon_{n\uparrow}$ with the YSR in-gap states displayed as red stars; (d,h) The spatial distribution of the quantity $|u_{r0\uparrow}|^2+|v_{r0\downarrow}|^2$ for the YSR state along the $c$ and $c'$ paths versus the distance to the impurity $r$.}
\label{fig2}
\end{figure*}

\begin{figure}[t]
\centering
\includegraphics[width=0.7\linewidth]{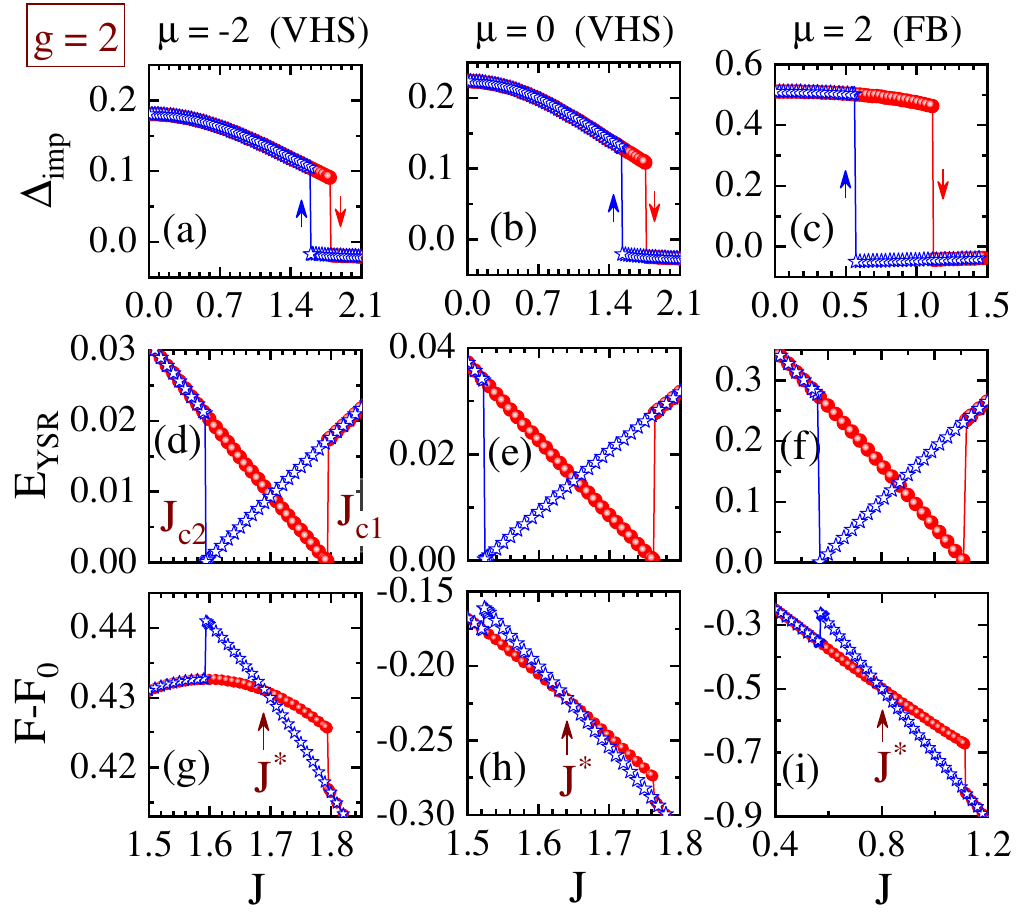}
\caption{(Color online) The pair potential at the impurity location $\Delta_{\rm imp}$ (a-c), the YSR state energy $E_{\rm YSR}$ (d-f), and the free energy $F$~(g-i) versus the coupling $J$ for $\mu=-2$ (VHS), $0$ (VHS) and $2$ (FB) at $g=2$ and $T=0$. The curves with red spheres and blue stars represent the ascending and descending branches, respectively, see the blue and red arrows shown in panels (a-c). $J_{\rm c1}$ and $J_{\rm c2}$ highlighted in panel (d) are the upper and lower critical interactions for $E_{\rm YSR}=0$. $J^*$ marks the point of the first-order transition in panels (g-i) and corresponds to the minimal free-energy-favorite (stable) YSR energy.
}\label{fig3}
\end{figure}

For $J$ near the first-order transition, one needs to capture the related hysteresis behavior. In this case, we apply a history-dependent initialization~\cite{chen2012a, shanenko2008a} in the ascending and descending manners. For the ascending/descending calculations, $J$ is increased/decreased by a sufficiently small step $\delta J$, and the BdG equations utilize the pair potential of the previous step as the initial condition. This makes it possible to get a hysteresis loop around the first-order phase transition when varying $J$, as seen in Figs.~\ref{fig2} and \ref{fig3}.

In our work, the energy-related quantities (e.g., $g$, $\mu$, $K$, $J$, $\Delta_i$ and $F$) are measured in units of the nearest-neighbor hopping parameter $t$. We consider a kagome lattice with $N\times N=20\times20$ primitive unit cells. The chosen value of $N$ is sufficiently large to reach the regime of an isolated YSR state. The doped magnetic impurity is positioned at the site $i_0=\{19, 20\}$ and marked by the red star in Fig.~\ref{fig1}(a). For our calculations, we adopt $K=0$ and $\hbar\omega_D=8$. Since the single-particle energy varies from $-4$ to $2$ in the non-superconducting kagome model, the adopted values of $\hbar\omega_D$ are large enough to include all the quasiparticle states with positive energies when calculating the pair potential given by Eq.~(\ref{OP}). This choice of $K$ and $\hbar \omega_D$ is not critical for our qualitative conclusions, other sets of these parameters produce similar results.

To have an idea about the impurity-free $s$-wave kagome model, the related results are demonstrated in Figs.~\ref{fig1}(b-f). In particular, Fig.~\ref{fig1}(b) shows self-consistent numerical results for the zero-temperature filling level $\overline{n}_e$ versus the chemical potential $\mu$~($J=0$). The curves with red spheres and green stars correspond to the pair couplings $g=2$ and $3$, respectively. The choice of $g=2$ makes it possible to get the pair potential similar to that of Ref.~{\cite{basak2022}}~(at $\mu=0$) while $g=3$ has been commonly used in previous theoretical works~\cite{tanaka2000, glodzik2018, bai2023,chen2024}. The black solid curve represents $\overline{n}_e$ for the normal system with $g=0$. One can see that $\overline{n}_e$ monotonously increases with $\mu$, as electrons fill the lattice. However, the slope of $\overline{n}_e$ as a function of $\mu$ is sensitive to the presence of the van Hove singularities, Dirac point, and flat band in the single-particle spectrum of the model. 

Self-consistent results for the temperature-dependent pair potential in the impurity-free case are given in Figs.~\ref{fig1}(c,d). The impurity-free system exhibits a spatially uniform condensate distribution $\Delta=\Delta_i$, and the value of $\Delta$ changes significantly with $\mu$~(its maximum corresponds to the FB regime). $\Delta$ calculated for $g=2$ at $\mu=0$ (VHS) and $2$ (FB) is shown in Fig.~\ref{fig1}(c), while Fig.~\ref{fig1}(d) demonstrates the temperature-dependent $\Delta$ for $g=3$ at $\mu=-1$ (DP), $0$ (VHS) and $2$ (FB). These results are similar to those of the self-consistent gap equation for conventional superconductors~\cite{ketterson1999}.

The impurity-free $\Delta$ as a function of $\mu$ is shown in Fig.~\ref{fig1}(e) for $T=0$, which can be compared with the effective density of states (effective DOS, defined as $d\overline{n}_e/d\mu$ at $T=0$ and given in units of $1/t$) in Fig.~\ref{fig1}(f). The dependence of the pair potential on $\mu$ is not that trivial. In particular, for $g=2$, the pair potential $\Delta$ exhibits three local maxima at the VHS and FB points, which are associated with the corresponding maxima in the effective DOS. The two VHS maxima are different because the VHS points $\mu=-2$ and $0$ correspond to different electron-filling levels. In addition, the superconductive correlations vanish around the DP in the interval $-1.3<\mu<-0.6$ (for $g=2$) due to a significant drop of the effective DOS to zero at the DP, see Fig.~\ref{fig1}(f). It is also worth mentioning that $\Delta$ is nonzero up to $\mu_{\rm max}=2.47$ which exceeds the chemical potential $\mu=2$ at the FB by about $24\%$. This feature appears due to a smearing of the FB in the presence of the superconductive correlations. This smearing is also visible in the electron filling level in Fig.~\ref{fig1}(b). The derivative of $\overline{n}_e$ with respect to $\mu$, i.e., the effective DOS in Fig.~\ref{fig1}(f), is no longer infinite at $\mu=2$ for $g=2$, and it drops to zero only at $\mu_{\rm max}=2.47$.

For $g=3$ (represented by the curves with green stars in Fig.~\ref{fig1}(e)), the pair potential is significantly enhanced as compared to the case of $g=2$. In more detail, the local maxima related to the VHS ($\mu=0$) and the FB merge in one pronounced peak of $\Delta$. In addition, a drop of $\Delta$ to zero near the DP observed at $g=2$, is replaced by a finite local minimum at $\mu=-1$ for $g=3$. The full occupation by electrons occurs at $\mu_{\rm max}=2.75$ which is larger than the corresponding maximum value of the chemical potential $2.47$ at $g=2$. One may notice that there is a distinctive shift between the peak positions of $\Delta$ and the effective DOS near the FB. We expect that this occurs due to the proximity of the VHS increase of $\Delta$ at $\mu=0$. 

Figure~\ref{fig2} shows two typical examples of the spatial distribution of the condensate and YSR state in the vicinity of the magnetic impurity, as calculated at $T=0$ and $\mu=0$~(VHS) for $g=2, J=2$~(the upper panels) and $g=3, J=2.4$~(the lower panels). Panels (a) and (e) illustrate the contour plots of the pair potential $\Delta_i$ near the magnetic impurity located at the intersection of lines $c$ and dense $c'$. One can see an overall depression of the pair potential near the impurity site. Additional details can be found in the spatial profiles of the pair potential $\Delta_r$ given as a function of the distance to the impurity along the horizontal $c'$ and slant $c$ directions, see Figs.~\ref{fig2}(b,f). Fast oscillations can be found in $\Delta_r$ along the slant direction $c$ while such oscillations disappear along the horizontal path $c'$. The healing length of this depression is estimated as $\sim 3a$, with $a$ the lattice constant.  

In Fig.~\ref{fig2}(c) and (g), we show the quasiparticle energy spectra corresponding to the parameters given in panels (a) and (e), respectively. The YSR states are marked as the red stars with $E_{\rm YSR} = 0.052$ and $0.164$, which are $23.3\%$ and $24.7\%$ of the impurity-free $\Delta$. 
The corresponding spatial profiles of $|u_{r0\uparrow}|^2+|v_{r0\downarrow}|^2$ for the YSR state are shown in Fig.~\ref{fig2}(d,h). [This quantity can be regarded as the squared modulus of the YSR wave function.] For the weaker coupling $g=2$, the YSR distribution along the slant (horizontal) path decays with (without) fast oscillations away from the impurity. For the stronger coupling $g=3$, the YSR state exhibits a monotonous spatial decay in both $c$ and $c'$ directions. The half-peak width of the YSR state in both cases is close to the healing length of the pair-potential depression, i.e. about $\sim 3a$. Thus, our lattice size with $20\times20$ primitive unit cells and the length $40a$ in both the horizontal and dense directions is sufficiently large to consider the response of the system to the presence of magnetic impurity.

Now we turn to the interplay of the classical magnetic impurity with the superconductive correlations. The related self-consistent results are shown in Fig.~\ref{fig3} for $g=2$ and $T=0$. In particular, the local pair potential $\Delta_{\rm imp}$ at the impurity location is shown versus the spin-magnetic coupling $J$ in Figs.~\ref{fig3}(a-c) for $\mu=-2$ (VHS), $0$ (VHS), and $2$ (FB). The energy of the YSR state $E_{\rm YSR}$ is given versus $J$ for the VHS, DP, and FB regimes in panels (d-f). Finally, the free energy $F$ as a function of $J$ is demonstrated in panels (g-i) for the same values of $\mu$. $F$ is shown relative to $F_0$, being the system free energy at $J=0$. To identify the presence of the first-order transition, the calculations are done in ascending and descending manners, using the pair potential distribution $\Delta_i$ calculated at the preceding step as the initial guess for the new step, corresponding to a new (larger or smaller) value of $J$. The curves with red spheres in Fig.~\ref{fig3} represent our results obtained in the ascending manner while the blue stars are calculated in the descending procedure. Notice that the range of $J$-values is deliberately enlarged in panels (d-i).

\begin{figure*}[ht]
\centering
\includegraphics[width=0.9\linewidth]{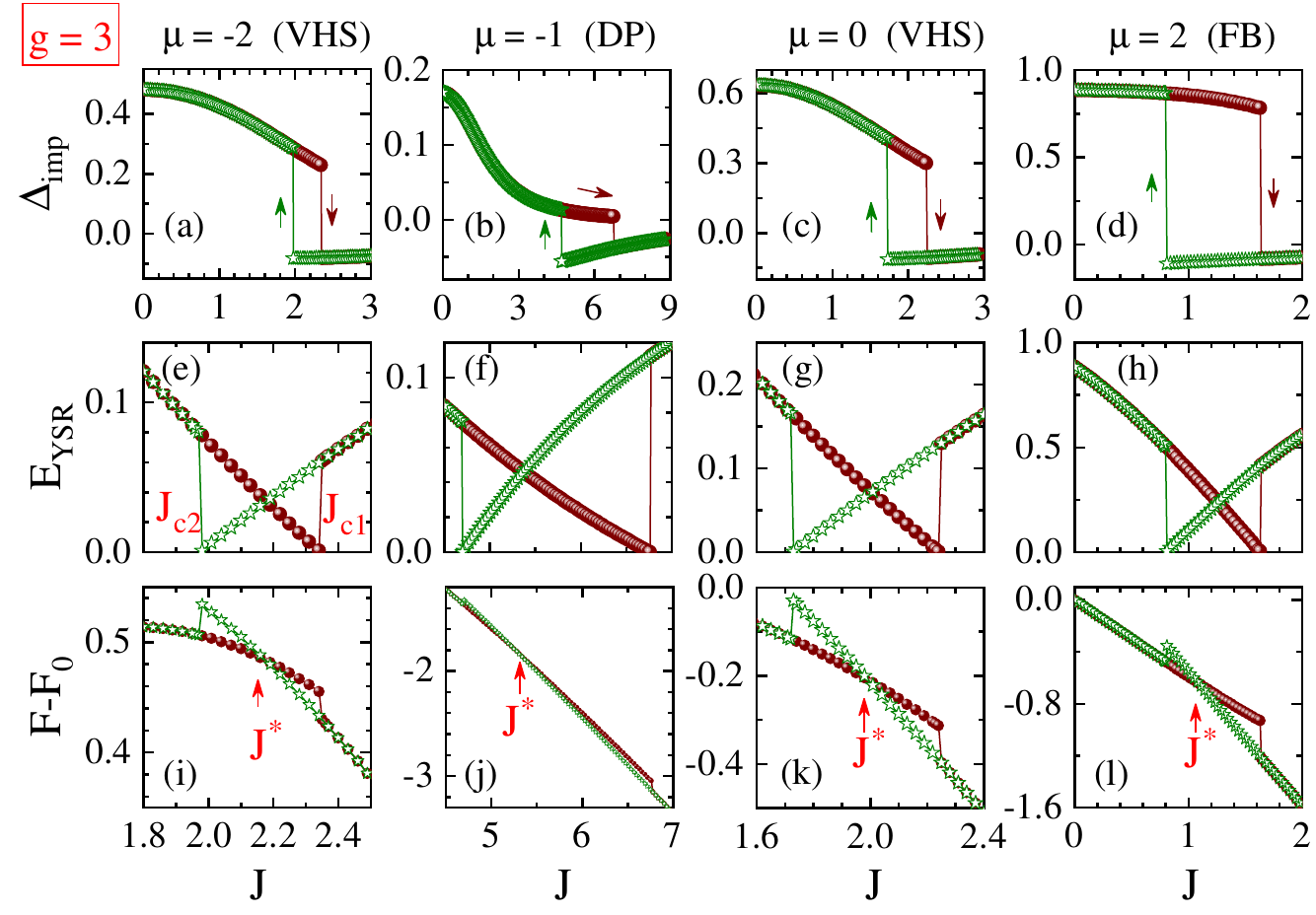}
\caption{(Color online) The same as in Fig.~\ref{fig3} but for $g=3$. In addition, Panels (b, f, j) are included with the results for $\mu=-1$ (DP). The ascending and descending branches are given by the curves with the wine-color spheres and green stars, respectively.} 
\label{fig4}
\end{figure*}

As compared to Ref.~\cite{basak2022}, the most distinctive feature in Fig.~\ref{fig3} is the presence of the first-order transition: one can see the discontinuities and the related hysteresis loops of $\Delta_{\rm imp}$, $E_{\rm YSR}$, and $F-F_0$. The jump in the ascending branch occurs at $J=J_{\rm c1}$ while the corresponding jump in the descending branch takes place at $J_{\rm c2} < J_{\rm c1}$. From Figs.~\ref{fig3}(a-c) one can learn that $\Delta_{\rm imp}$ is positive for both the ascending and descending branches for $J < J_{\rm c1}$ and $J < J_{\rm c2}$, respectively. However, irrespective of the particular branch, $\Delta_{\rm imp}$ becomes negative above the corresponding jump point. This is similar to the $0$-$\pi$ transition in the phase of the pair potential reported previously for the case of a magnetic impurity in the square and triangular lattices~\cite{salkola1997, glodzik2018}. One can also see that below the transition region, $\Delta_{\rm imp}$ decreases with increasing $J$, as seen for $\mu=-2$, $0$, and $2$ in Figs.~\ref{fig3}(a-c). This agrees with the earlier results of Refs.~\cite{balatsky2006,salkola1997}. 

From Fig.~\ref{fig3}(d-f), we find that $E_{\rm YSR}=0$ at $J=J_{\rm c1}$ and $J=J_{\rm c2}$, which differs significantly from the result $J_{\rm c}=1/\pi D(\mu)$~(with $D(\mu)$ the DOS per spin at $\mu$) obtained in Refs.~\cite{balatsky2006, basak2022} for $K=0$ within the assumption of a constant pair potential. Our numerical study reveals that $J_{\rm c1}$ and $J_{\rm c2}$ are very sensitive to $\mu$. For $J_{\rm c1}$ we get $1.79$~($\mu = -2$, VHS), $1.76$~($\mu=0$, VHS), and $1.11$~($\mu=2$, FB). For $J_{\rm c2}$ one obtains $1.59$~($\mu = -2$, VHS), $1.52$~($\mu=0$, VHS), and $0.57$~($\mu=2$, FB). Moreover, the difference $J_{\rm c1}-J_{\rm c2}$ also depends strongly on $\mu$: $J_{\rm c1}-J_{\rm c2} = 0.20$, $0.24$, and $0.54$ for $\mu = -2$, $0$, and $2$, respectively. It is instructive to compare the critical value $J_{\rm c}=1/\pi D(\mu)$ with our numerical results for $J_{\rm c1}$ and $J_{\rm c2}$. To estimate $D(\mu)$, one can employ the effective DOS of our kagome model in Fig.~\ref{fig1}(f). The estimate results in $J_c = 1.63$, $1.71$, and $0.83$ for $\mu=-2$, $0$ and $2$, respectively. We obviously have $J_{\rm c1}< J_{\rm c} < J_{\rm c2}$.  

To determine the stable YSR states, one needs to compare the free energies of the ascending (red spheres) and descending (blue stars) branches. Then, according to the standard prescriptions of thermodynamics, the stable YSR states correspond to the minimal free energy. One can see from Figs.~\ref{fig3}(g-i) that the free energies of the ascending and descending branches cross each other at $J^*$, which is the point of the first-order transition occurring in the system with changing the impurity-magnetic interaction $J$. At the same time, according to Figs.~\ref{fig3}(a-c), this is the $0$-$\pi$ transition in the phase of the pair potential that also appears for a magnetic impurity in square and triangular lattices~\cite{salkola1997, glodzik2018}. For $\mu=-2$, $0$ and $2$, one finds $J^*=1.69$, $1.64$ and $0.80$, as seen from Figs.~\ref{fig3}(g-i). Notice that these values are close to the estimates of $J_c=1/\pi D(\mu)$ found in the previous paragraph. However, contrary to the calculations~\cite{balatsky2006, basak2022} assuming a constant pair potential, the stable YSR states have nonzero energies $E_{\rm YSR} \not=0$.

To demonstrate that our general conclusions are not sensitive to a particular value of the pair coupling $g$, Fig.~\ref{fig4} illustrates the results calculated at $g=3$~($T=0$) for $\mu = -2$ (VHS), $-1$ (DP), $0$ (VHS), and $2$ (FB). In addition to the values of $\mu$ given in Fig.~\ref{fig3}, now we have also $\mu=-1$ corresponding to the DP. At $g=2$ the system is in the normal state near this point while at $g=3$ the corresponding pair potential remains finite. Here, the branches calculated in the ascending and descending manners are marked with wine-color spheres and green stars, respectively. Similarly to the results in Fig.~\ref{fig3}, our study again reveals the first-order transition, i.e., the discontinuities of $\Delta_{\rm imp}$, $E_{\rm YSR}$ and $F-F_0$ as functions of $J$ are clearly seen in Fig.~\ref{fig4}. One notes that the decrease of $\Delta_{\rm imp}$ with increasing $J$ below the transition is much more pronounced for the DP in comparison with the other choices of $\mu$, see Fig.~\ref{fig4}(b). In addition, $\Delta_{\rm imp}$ shown in Figs.~\ref{fig4}(a,c,d) for $\mu=-2$, $0$, and $2$, is notably larger than its counterpart in Figs.~\ref{fig3}(a-c) due to the increase of the coupling $g$. 

Comparing the results for $g=3$ and $g=2$, we find that $J_{\rm c1}$ and $J_{\rm c2}$ depend significantly on $g$. One can see from Fig.~\ref{fig4} that for $\mu=-2$, $0$ and $2$ we have $J_{\rm c1} = 2.34$, $2.24$, $1.63$~($g=2$) and $J_{\rm c2} = 1.98$, $1.73$ and $0.81$~($g=3$). The estimates of $J_{\rm c}$ utilizing the effective DOS of Fig.~\ref{fig1}(f), yield $2.21$, $2.00$, and $1.06$ for the above values of $\mu$. For these estimates we again find the inequality $J_{\rm c2} < J_{\rm c} < J_{\rm c1}$. However, for $\mu = -1$ (DP), this inequality does not hold. Instead, we have $J_{\rm c1} = 6.76$, $J_{\rm c2} = 4.70$ while $J_c=12.48$, i.e., $J_{\rm c} > J_{\rm c1,c2}$. 
This is due to the fact that $D(\mu)=0$ at the Dirac point [as seen in Fig.~\ref{fig1}(f)], which leads to an infinity $J_c[= 1/\pi D(\mu)]$. Finally, for all these four values of $\mu$, the free energies of both the ascending and descending branches cross each other at the first-order transition point $J^*$, although these branches go very close to one another in the case of $\mu=1$. Therefore, the energies of the stable YSR states can not reach zero, which agrees with the results for $g=2$ in Fig.~\ref{fig3}.

\begin{figure}[ht]
\centering
\includegraphics[width=0.5\linewidth]{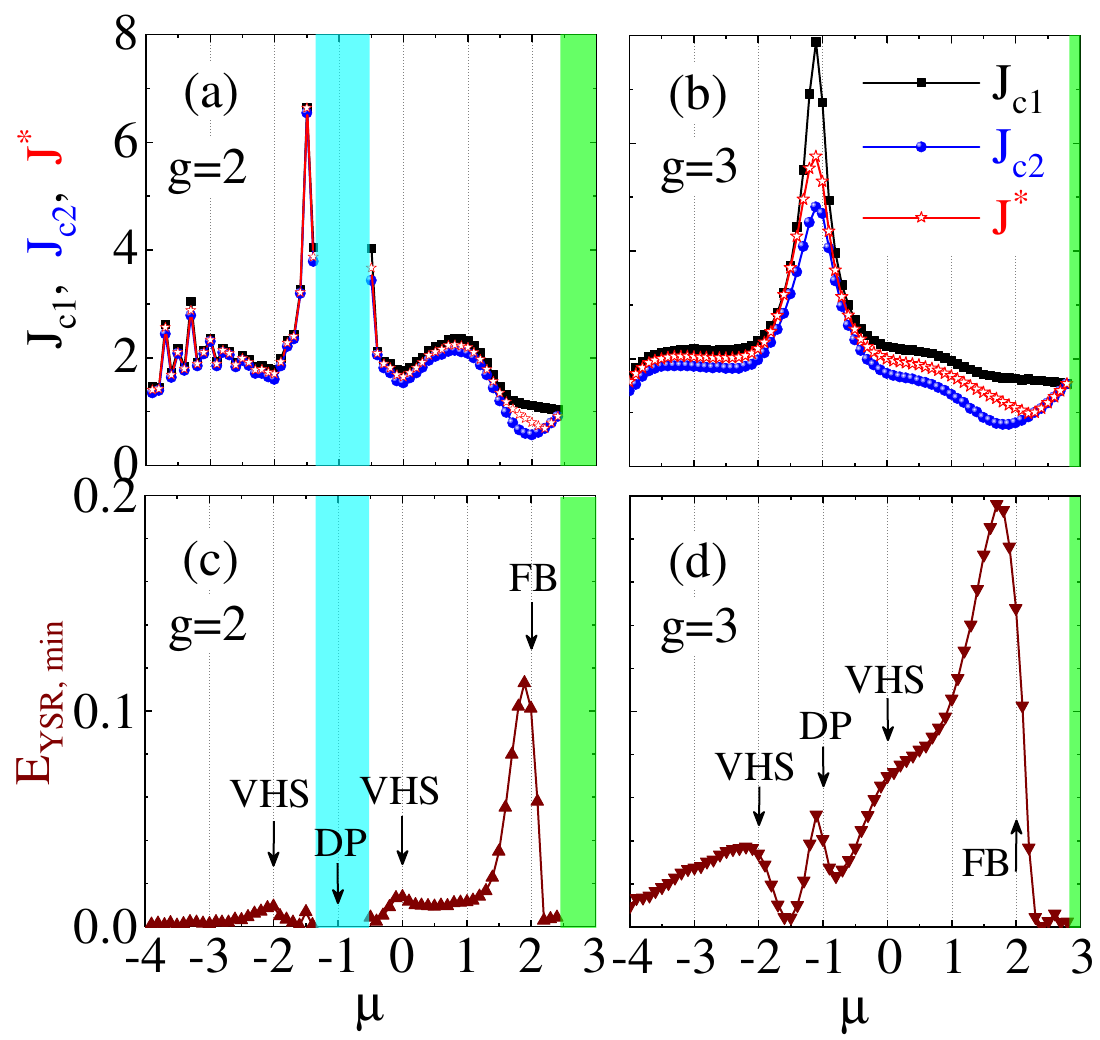}
\caption{(Color online) The first-order transition characteristics $J_{\rm c1}$, $J_{\rm c2}$, and $J^*$~(a, b) together with $E_{\rm YSR, min}$~(c, d) are given as functions of $\mu$ for $g=2$ and $3$ at $T=0$. The areas marked in green on the right side of all the panels, highlight the electron full-filling regime. The blue shaded regions in panels (a, c) correspond to the suppression of the superconductive correlations due to approaching the DP regime at $g=2$, Fig.~\ref{fig1}(e).}
\label{fig5}
\end{figure}

Now, to go into more detail concerning the first-order phase transition, we consider how its characteristics depend on $\mu$. In particular, $J_{\rm c1}$~(curves with black squares), $J_{\rm c2}$~(curves with blue spheres), and the first-order transition point $J^*$~(curves with red stars) are shown versus $\mu$ in Figs.~\ref{fig5}(a,b) for $g=2$ and $3$. In addition, the corresponding minimal energy of the stable YSR states $E_{\rm YSR, min}$ is given as a function of $\mu$ for the same couplings in Figs.~\ref{fig5}(c, d). The other calculation parameters are the same as in Figs.~\ref{fig1}$-$\ref{fig4}. The blue and green regions in these plots highlight the complete suppression of the superconductive correlations near the DP (for $g=2$) and within the full-filling regime, respectively.

The general profiles of $J_{\rm c1}$, $J_{\rm c2}$, and $J^*$ in Fig.~\ref{fig5}(a), corresponding to $g=2$, look similar to the dependence of $J_c$ on $\mu$ reported in Fig. 2(b) of Ref.~\cite{basak2022} but with two exceptions: the first one concerns the absence of the superconductive correlations near the DP~(the blue region); the second exception is that the onset of the full-filling regime $\mu_{\rm max}$ is larger than $\mu=2$ due to a smearing of the FB caused by the superconducting correlations within the self-consistent calculations. However, despite this similarity, the physics behind our calculations differs significantly from that of the non-self-consistent study assuming a constant pair potential. The latter yields~\cite{balatsky2006, basak2022} that when changing the interaction strength $J$, there is a crossover at the point $J=J_{\rm c}$ with zero corresponding YSR energy. Our self-consistent consideration demonstrates that the zero-energy YSR states are metastable and appear in the ascending and descending branches of the first-order-transition hysteresis loop, see Figs.~\ref{fig1} and \ref{fig3}. The minimal stable YSR energy is always positive (though it can be very small) and corresponds to the point of the first-order phase transition $J^*$, associated with the $0$-$\pi$ transition of the pair potential and located between $J_{\rm c2}$ and $J_{\rm c1}$. As seen from Fig.~\ref{fig5}(a), the size of the hysteresis loop (the difference between $J_{\rm c1}$ and $J_{\rm c2}$) increases with $\mu$ and reaches its maximum near $\mu=2$, about the smeared FB. The corresponding $E_{\rm YSR, min}$ shown in Fig.~\ref{fig5}(c), has its global maximum $0.11$ at $\mu=1.9$~(near the FB) together with the two local maxima at the smeared VHS points near $\mu=-2$ and $0$. 

The results for the $\mu$-dependence of $J_{\rm c1}$, $J_{\rm c2}$, and $J^*$ for $g=3$ are shown in Fig.~\ref{fig5}(b). They are qualitatively similar to the results in Fig.~\ref{fig5}(a) and differ from them only due to the absence of the DP suppression of the superconductive state and the increase of the onset of the full-filling regime to $\mu_{\rm max}=2.75$. Now the size of the hysteresis loop $J_{c1}-J_{c2}$ exhibits an increase not only near the FB~($\mu=2$) but also at the DP~(about $\mu=-1$). From Fig.~\ref{fig5}(d) one can learn that $E_{\rm YSR, min}$ has two local minima at $\mu=-2.2$ and $-1.1$ accompanied by the maxima connected with the VHS, DP, and the FB. The two maxima of $E_{\rm YSR, min}$ visible at $\mu=0$ and $1.9$ for $g=2$ in Fig.~\ref{fig5}(c), merge into one profound increase of $E_{\rm YSR, min}$ at $\mu=1.8$ in Fig.~\ref{fig5}(d). In general, at any $\mu$ and $g$ our study reveals the first-order phase transitions associated with the dependence of the pertinent quantities on $J$. Irrespective of the superconductive coupling $g$, the minimal stable YSR energy is nonzero and demonstrates a significant increase near the VHS, DP, and FB. We stress that this feature cannot be simply related to an increase of the pair potential: at the DP (see the data for $g=3$) the pair potential goes down. 

In conclusion, based on a self-consistent numerical solution of the Bogoliubov-de Gennes equations, we have investigated the Yu-Shiba-Rusinov (YSR) states in the $s-$wave superconducting kagome model. Following the density of states, the bulk superconducting pair potential changes with the chemical potential $\mu$, reaching local maxima near the van Hove singularities (VHS) and the flat band (FB) while exhibiting the local minimum at the Dirac point (DP). 

When varying the strength $J$ of the interaction between a magnetic impurity and superconductor, the first-order phase transition occurs, accompanied by the pronounced hysteresis loop. The stable branch (with minimal free energy) exhibits the $0$-$\pi$ phase shift of the pair potential (abrupt change of the potential sign) at the transition point $J=J^*$. The corresponding YSR energy $E_{\rm YSR}$ has a jump at $J^*$ as well. The minimal $E_{\rm YSR}$ of the stable YSR states taken as a function of $J$ is positive for any set of the model parameters, i.e., $E_{\rm YSR, min} >0$. 

Considered as a function of $\mu$, $E_{\rm YSR, min}$ has three local maxima near the VHS, DP, and FB. The most pronounced increase of $E_{\rm YSR, min}$ is observed in the FB regime. For a shallow system, just above $\mu=-4$, and close to the full-filling onset, $E_{\rm YSR, min}$ drops significantly but remains nonzero. The first-order transition characteristics $J_{\rm c1}$, $J_{\rm c2}$, and $J^*$ depend on $g$ but the qualitative features of their behavior do not change with $g$. Finally, we expect that our findings can be helpful for future experiments on the YSR states in the $s-$wave superconducting kagome systems achieved by the superconducting proximity effects.

\section*{Acknowledgements}
This work was supported by the Science Foundation of Zhejiang Sci-Tech University (ZSTU) (Grants No. 19062463-Y) and funded within the framework of the HSE University Basic Research Program.


\section*{References}
\begin{thebibliography}{100}
\bibitem{guo2009} Guo, H.-M.; Franz, M. \href{https://doi.org/10.1103/PhysRevB.80.113102}{Topological insulator on the kagome lattice}, \textit{Phys. Rev. B} \textbf{2009}, \textit{80}, 113102.
\bibitem{ko2009} Ko, W.-H.; Lee, P.~A.; Wen, X.-G. \href{https://doi.org/10.1103/PhysRevB.79.214502}{Doped kagome system as exotic superconductor}, \textit{Phys. Rev. B} \textbf{2009}, \textit{79}, 214502.
\bibitem{kiesel2012} Kiesel, M.~L.; Thomale, R. \href{https://doi.org/10.1103/PhysRevB.86.121105}{Sublattice interference in the kagome Hubbard model}, \textit{Phys. Rev. B} \textbf{2012}, \textit{86}, 121105(R).
\bibitem{wang2013c} Wang, W.-S.; Li, Z.-Z.; Xiang, Y.-Y.; Wang, Q.-H. \href{https://doi.org/10.1103/PhysRevB.87.115135}{Competing electronic orders on kagome lattices at van Hove filling}, \textit{Phys. Rev. B} \textbf{2013}, \textit{87}, 115135.
\bibitem{ye2018a} Ye, L.; Kang, M.; Liu, J.; von Cube, F.; Wicker, C.~R.; Suzuki, T.; Jozwiak, C.; Bostwick, A.; Rotenberg, E.; Bell, D.~C.; Fu, L.; Comin, R.; Checkelsky, J.~G. \href{https://doi.org/10.1038/nature25987}{Massive Dirac fermions in a ferromagnetic kagome metal}, \textit{Nature} \textbf{2018}, \textit{555}, 638.
\bibitem{ortiz2019} Ortiz, B.~R.; Gomes, L.~C.; Morey, J.~R.; Winiarski, M.; Bordelon, M.; Mangum, J.~S.; Oswald, I.~W.~H.; Rodriguez-Rivera, J.~A.; Neilson, J.~R.; Wilson, S.~D.; Ertekin, E.; McQueen, T.~M.; Toberer, E.~S. \href{https://doi.org/10.1103/PhysRevMaterials.3.094407}{New kagome prototype materials: Discovery of KV$_3$Sb$_5$, RbV$_3$Sb$_5$, and CsV$_3$Sb$_5$}, \textit{Phys. Rev. Materials} \textbf{2019}, \textit{3}, 094407.
\bibitem{ortiz2020} Ortiz, B.~R.; Teicher, S.~M.~L.; Hu, Y.; Zuo, J.~L.; Sarte, P.~M.; Schueller, E.~C.; Abeykoon, A.~M.~M.; Krogstad, M.~J.; Rosenkranz, S.; Osborn, R.; Seshadri, R.; Balents, L.; He, J.; Wilson, S.~D. \href{https://doi.org/10.1103/PhysRevLett.125.247002}{CsV$_3$Sb$_5$: A Z$_2$ Topological Kagome Metal with a Superconducting Ground State}, \textit{Phys. Rev. Lett.} \textbf{2020}, \textit{125}, 247002.
\bibitem{hu2022a} Hu, Y.; Wu, X.; Ortiz, B.~R.; Ju, S.; Han, X.; Ma, J.; Plumb, N.~C.; Radovic, M.; Thomale, R.; Wilson, S.~D.; Schnyder, A.~P.; Shi, M. \href{https://doi.org/10.1038/s41467-022-29828-x}{Rich nature of Van Hove singularities in Kagome superconductor CsV$_3$Sb$_5$}, \textit{Nat. Commun.} \textbf{2022}, \textit{13}, 2220.
\bibitem{wang2020b} Wang, P.; Wang, Y.; Zhang, B.; Li, Y.; Wang, S.; Wu, Y.; Zhu, H.; Liu, Y.; Zhang, G.; Liu, D.; Xiong, Y.; Sun, Z. \href{https://doi.org/10.1088/0256-307X/37/8/087102}{Experimental Observation of Electronic Structures of Kagome Metal YCr$_6$Ge$_6$}, \textit{Chinese Phys. Lett.} \textbf{2020}, \textit{37}, 087102.
\bibitem{pokharel2021} Pokharel, G.; Teicher, S.~M.~L.; Ortiz, B.~R.; Sarte, P.~M.; Wu, G.; Peng, S.; He, J.; Seshadri, R.; Wilson, S.~D. \href{https://doi.org/10.1103/PhysRevB.104.235139}{Electronic properties of the topological kagome metals YV$_6$Sn$_6$ and GdV$_6$Sn$_6$}, \textit{Phys. Rev. B} \textbf{2021}, \textit{104}, 235139.
\bibitem{chen2021c} Chen, K.~Y.; Wang, N.~N.; Yin, Q.~W.; Gu, Y.~H.; Jiang, K.; Tu, Z.~J.; Gong, C.~S.; Uwatoko, Y.; Sun, J.~P.; Lei, H.~C.; Hu, J.~P.; Cheng, J.-G. \href{https://doi.org/10.1103/PhysRevLett.126.247001}{Double Superconducting Dome and Triple Enhancement of $T_c$ in the Kagome Superconductor CsV$_3$Sb$_5$ under High Pressure}, \textit{Phys. Rev. Lett.} \textbf{2021}, \textit{126}, 247001.
\bibitem{mielke2022} Mielke, C.; Das, D.; Yin, J.-X.; Liu, H.; Gupta, R.; Jiang, Y.-X.; Medarde, M.; Wu, X.; Lei, H.~C.; Chang, J.; Dai, P.; Si, Q.; Miao, H.; Thomale, R.; Neupert, T.; Shi, Y.; Khasanov, R.; Hasan, M.~Z.; Luetkens, H.; Guguchia, Z. \href{https://doi.org/10.1038/s41586-021-04327-z}{Time-reversal symmetry-breaking charge order in a kagome superconductor}, \textit{Nature} \textbf{2022}, \textit{602}, 245.
\bibitem{nie2022} Nie, L.; Sun, K.; Ma, W.; Song, D.; Zheng, L.; Liang, Z.; Wu, P.; Yu, F.; Li, J.; Shan, M.; Zhao, D.; Li, S.; Kang, B.; Wu, Z.; Zhou, Y.; Liu, K.; Xiang, Z.; Ying, J.; Wang, Z.; Wu, T.; Chen, X. \href{https://doi.org/10.1038/s41586-022-04493-8}{Charge-density-wave-driven electronic nematicity in a kagome superconductor}, \textit{Nature} \textbf{2022}, \textit{604}, 59.
\bibitem{chen2021} Chen, H.; Yang, H.; Hu, B.; Zhao, Z.; Yuan, J.; Xing, Y.; Qian, G.; Huang, Z.; Li, G.; Ye, Y.; Ma, S.; Ni, S.; Zhang, H.; Yin, Q.; Gong, C.; Tu, Z.; Lei, H.; Tan, H.; Zhou, S.; Shen, C.; Dong, X.; Yan, B.; Wang, Z.; Gao, H.-J. \href{https://doi.org/10.1038/s41586-021-03983-5}{Roton pair density wave in a strong-coupling kagome superconductor}, \textit{Nature} \textbf{2021}, \textit{599}, 222.
\bibitem{yu2021} Yu, F.~H.; Ma, D.~H.; Zhuo, W.~Z.; Liu, S.~Q.; Wen, X.~K.; Lei, B.; Ying, J.~J.; Chen, X.~H. \href{https://doi.org/10.1038/s41467-021-23928-w}{Unusual competition of superconductivity and charge-density-wave state in a compressed topological kagome metal}, \textit{Nat. Commun.} \textbf{2021}, \textit{12}, 3645.
\bibitem{xu2021} Xu, H.-S.; Yan, Y.-J.; Yin, R.; Xia, W.; Fang, S.; Chen, Z.; Li, Y.; Yang, W.; Guo, Y.; Feng, D.-L. \href{https://doi.org/10.1103/PhysRevLett.127.187004}{Multiband Superconductivity with Sign-Preserving Order Parameter in Kagome Superconductor CsV$_3$Sb$_5$}, \textit{Phys. Rev. Lett.} \textbf{2021}, \textit{127}, 187004.
\bibitem{gupta2022} Gupta, R.; Das, D.; Mielke~III, C.~H.; Guguchia, Z.; Shiroka, T.; Baines, C.; Bartkowiak, M.; Luetkens, H.; Khasanov, R.; Yin, Q.; Tu, Z.; Gong, C.; Lei, H. \href{https://doi.org/10.1038/s41535-022-00453-7}{Microscopic evidence for anisotropic multigap superconductivity in the CsV$_3$Sb$_5$ kagome superconductor}, \textit{npj Quantum Mater.} \textbf{2022}, \textit{7}, 49.
\bibitem{yu1965} Yu, L. Bound State in Superconductors With Paramagnetic Impurities, \textit{Acta Phys. Sin.} \textbf{1965}, \textit{21}, 75.
\bibitem{shiba1968} Shiba, H. Classical Spins in Superconductors, \textit{Prog. Theor. Phys.} \textbf{1968}, \textit{40}, 435.
\bibitem{rusinov1969a} Rusinov, A.~I. On the theory of gapless superconductivity in alloys containing paramagnetic impurities, \textit{Sov. Phys. JETP} \textbf{1969}, \textit{29}, 1101.
\bibitem{balatsky2006} Balatsky, A.~V.; Vekhter, I.; Zhu, J.-X. \href{https://doi.org/10.1103/RevModPhys.78.373}{Impurity-induced states in conventional and unconventional superconductors}, \textit{Rev. Mod. Phys.} \textbf{2006}, \textit{78}, 373.
\bibitem{Kitaev2001} Kitaev, A.~Y. \href{https://doi.org/10.1070/1063-7869/44/10S/S29}{Unpaired Majorana fermions in quantum wires}, \textit{Phys.-Usp.} \textbf{2001}, \textit{44}, 131.
\bibitem{morr2003} Morr, D.~K.; Stavropoulos, N.~A. \href{https://doi.org/10.1103/PhysRevB.67.020502}{Quantum interference between impurities: Creating novel many-body states in $s-$wave superconductors}, \textit{Phys. Rev. B} \textbf{2003}, \textit{67}, 020502(R).
\bibitem{choy2011} Choy, T.-P.; Edge, J.~M.; Akhmerov, A.~R.; Beenakker, C.~W.~J. \href{https://doi.org/10.1103/PhysRevB.84.195442}{Majorana fermions emerging from magnetic nanoparticles on a superconductor without spin-orbit coupling}, \textit{Phys. Rev. B} \textbf{2011}, \textit{84}, 195442.
\bibitem{nadj-perge2013} Nadj-Perge, S.; Drozdov, I.~K.; Bernevig, B.~A.; Yazdani, A. \href{https://doi.org/10.1103/PhysRevB.88.020407}{Proposal for realizing Majorana fermions in chains of magnetic atoms on a superconductor}, \textit{Phys. Rev. B} \textbf{2013}, \textit{88}, 020407(R).
\bibitem{klinovaja2013} Klinovaja, J.; Stano, P.; Yazdani, A.; Loss, D. \href{https://doi.org/10.1103/PhysRevLett.111.186805}{Topological Superconductivity and Majorana Fermions in RKKY Systems}, \textit{Phys. Rev. Lett.} \textbf{2013}, \textit{111}, 186805.
\bibitem{basak2022} Basak, S.; Ptok, A. \href{https://doi.org/10.1103/PhysRevB.105.094204}{Shiba states in systems with density of states singularities}, \textit{Phys. Rev. B} \textbf{2022}, \textit{105}, 094204.
\bibitem{lin2022} Lin, Y.-H.; Chen, C.-J.; Kumar, N.; Yeh, T.-Y.; Lin, T.-H.; Bl\"{u}gel, S.; Bihlmayer, G.; Hsu, P.-J. \href{https://doi.org/10.1021/acs.nanolett.2c02831}{Fabrication and Imaging Monatomic Ni Kagome Lattice on Superconducting Pb(111)}, \textit{Nano Lett.} \textbf{2022}, \textit{22}, 8475.
\bibitem{farinacci2023} Farinacci, L.; Reecht, G.; von Oppen, F.; Franke, K. J. \href{https://arxiv.org/abs/2307.09993}{Yu-Shiba-Rusinov bands in a self-assembled kagome lattice of magnetic molecules}, \textit{arxiv} \textbf{2023}, 2307.09993.
\bibitem{salkola1997} Salkola, M.~I.; Balatsky, A.~V.; Schrieffer, J.~R. \href{https://doi.org/10.1103/PhysRevB.55.12648}{Spectral properties of quasiparticle excitations induced by magnetic moments in superconductors}, \textit{Phys. Rev. B} \textbf{1997}, \textit{55}, 12648.
\bibitem{glodzik2018} G\l{}odzik, S.; Ptok, A. \href{https://link.springer.com/article/10.1007/s10948-017-4360-6}{Quantum Phase Transition Induced by Magnetic Impurity: Triangular Lattice with On-Site Pairing Study}, \textit{J. Supercond. Nov. Magn.} \textbf{2018}, \textit{31}, 647.
\bibitem{Essler2005} Essler, F.~H.~L.; Frahm, H.; Gohmann, F.; Klumper, A.; Korepin, V.~E. \textit{The One-dimensional Hubbard Model} (Cambridge University Press, 2005).
\bibitem{correction} We have verified that the sign before $(K-\sigma J)\delta_{i0}$ should be a minus, which is different from Eq.~(7) in Ref.~\cite{basak2022}.
\bibitem{degennes1964} de~Gennes, P.~G. \href{https://doi.org/10.1103/RevModPhys.36.225}{Boundary Effects in Superconductors}, \textit{Rev. Mod. Phys.} \textbf{1964}, \textit{36}, 225.
\bibitem{ketterson1999} Ketterson, J.~B.; Song, S.~N. \textit{Superconductivity} (Cambridge, 1999).
\bibitem{chen2009} Chen, Y.; Croitoru, M.~D.; Shanenko, A.~A.; Peeters, F.~M. \href{https://doi.org/http://dx.doi.org/10.1088/0953-8984/21/43/435701}{Superconducting nanowires: Quantum confinement and spatially dependent Hartree-Fock potential}, \textit{J. Phys. Condens. Matter} \textbf{2009}, \textit{21}, 435701.
\bibitem{chen2012a} Chen, Y.; Shanenko, A.~A.; Croitoru, M.~D.; Peeters, F.~M. \href{http://dx.doi.org/10.1088/0953-8984/24/26/265702}{Quantum cascades in nano-engineered superconductors: Geometrical, thermal and paramagnetic effects}, \textit{J. Phys. Condens. Matter} \textbf{2012}, \textit{24}, 265702.
\bibitem{chen2014} Chen, Y.; Shanenko, A.~A.; Peeters, F.~M. \href{https://doi.org/10.1103/PhysRevB.89.054513}{Vortex anomaly in low-dimensional fermionic condensates: Quantum confinement breaks chirality}, \textit{Phys. Rev. B} \textbf{2014}, \textit{89}, 054513.
\bibitem{yin2023} Yin, L.; Bai, Y.; Zhang, M.; Shanenko, A.~A.; Chen, Y. \href{https://doi.org/10.1103/PhysRevB.108.054508}{Surface superconductor-insulator transition induced by electric field}, \textit{Phys. Rev. B} \textbf{2023}, \textit{108}, 054508.
\bibitem{chen2024} Chen, Y.; Zhu, Q.; Zhang, M.; Luo, X.; Shanenko, A.~A. \href{https://doi.org/10.1016/j.physleta.2023.129281}{Surface superconductor-insulator transition: Reduction of the critical electric field by Hartree-Fock potential}, \textit{Physics Letters A} \textbf{2024}, \textbf{494}, 129281.
\bibitem{kosztin1998a} Kosztin, I.; Kos, \v{S}.; Stone, M.; Leggett, A.~J. \href{https://doi.org/10.1103/PhysRevB.58.9365}{Free energy of an inhomogeneous superconductor: A wave-function approach}, \textit{Phys. Rev. B} \textbf{1998}, \textit{58}, 9365.
\bibitem{bai2023} Bai, Y.; Chen, Y.; Croitoru, M.~D.; Shanenko, A.~A.; Luo, X.; Zhang, Y. \href{https://doi.org/10.1103/PhysRevB.107.024510}{Interference-induced surface superconductivity: Enhancement by tuning the Debye energy}, \textit{Phys. Rev. B} \textbf{2023}, \textit{107}, 024510.
\bibitem{bai2023a} Bai, Y.; Zhang, L.; Luo, X.; Shanenko, A.~A.; Chen, Y. \href{https://doi.org/10.1103/PhysRevB.108.134506}{Tailoring of interference-induced surface superconductivity by an applied electric field}, \textit{Phys. Rev. B} \textbf{2023}, \textit{108}, 134506.

\bibitem{shanenko2008a} Shanenko, A. A.; Croitoru, M. D.; Peeters, F. M. \href{http://dx.doi.org/10.1103/PhysRevB.78.024505}{Magnetic-field induced quantum-size cascades in superconducting nanowires}, \textit{Phys. Rev. B} \textbf{2008}, \textit{78}, 024505.

\bibitem{tanaka2000} Tanaka, K.; Marsiglio, F. \href{http://dx.doi.org/10.1103/PhysRevB.62.5345}{Anderson prescription for surfaces and impurities}, \textit{Phys. Rev. B} \textbf{2000}, \textit{62}, 5345.
\end{thebibliography}
\end{document}